\documentclass[12pt,a4paper]{JHEP3}
\usepackage[all]{xy}
\usepackage{graphics}
\usepackage{epsfig}
\usepackage{amsmath}

\def\p{\partial}

\def\half{{1\over 2}}

\def\r{{\rho}}

\def\a{\alpha}
\def\b{\beta}
\def\g{\gamma}
\def\d{\delta}
\def\e{\epsilon}
\def\k{\kappa}
\def\m{\mu}

\def\n{\nu}

\def\r{\rho}
\def\s{\sigma}
\def\p{\partial}
\def\t{{\theta}}

\def\S{\Sigma}

\def\O{\Omega}
\def\G{\Gamma}
\def\D{\Delta}

\newcommand{\be}{\begin{equation}}
\newcommand{\ee}{\end{equation}}
\newcommand{\bea}{\begin{eqnarray}}
\newcommand{\eea}{\end{eqnarray}}
\newcommand{\Str}{\ensuremath{\mathrm{Str}}}

\title{Integrability of Green-Schwarz Sigma Models with Boundaries}

\author{Amit~Dekel and Yaron~Oz \\
  Raymond and Beverly Sackler School of Physics and Astronomy \\
  Tel-Aviv University, Ramat-Aviv 69978, Israel\\
\email{amitde@post.tau.ac.il}, \email{yaronoz@post.tau.ac.il}}

\abstract{
We construct integrability preserving boundary conditions for Green-Schwarz sigma-models on semi-symmetric spaces.
The boundary conditions are expressed as gluing conditions of the flat-connection, using an involutive metric preserving automorphism.
We show that the boundary conditions preserve half of the space-time supersymmetry and an infinite set of conserved charges.
We find integrable D-brane configurations for $\mathrm{AdS}_5\times \mathrm{S}^5$ and $\mathrm{AdS}_4\times \mathbb{C}\mathrm{P}^3$ backgrounds.
}


\preprint{}

\begin{document}

\section{Introduction}

Green-Schwarz sigma models (GS$\s$M) on the semi-symmetric spaces $\frac{\mathrm{PSU}(2,2|4)}{\mathrm{SO}(4,1)\times \mathrm{SO}(5)}\supset\mathrm{AdS}_5\times \mathrm{S}^5$ \cite{Metsaev:1998it} is known to be classically integrable \cite{Bena:2003wd}.
Similarly, any such GS$\s$M on semi-symmetric space is classically integrable as well \cite{Adam:2007ws}.
The existence of the infinite set of conserved charges depends on the boundary conditions of the string.
Usually in this context, the boundary conditions are taken to be either periodic, or the string's length being taken to infinity, where suitable boundary conditions are fixed.
When considering finite open strings,  the quantities that are conserved for closed strings are in general no longer conserved, since the monodromy matrix that generates the charges is not conserved.

In some cases, an infinite set of conserved charges can be generated from of an object that is constructed using the transition matrix and reflection matrices, and the model remains integrable.
These constructions were introduced first by Sklyanin in \cite{Sklyanin:1987bi}\cite{Sklyanin:1988yz}, and were used
to exhibit integrability of the affine Toda theory \cite{Bowcock:1995vp} and the O$(N)$ non-linear sigma-model \cite{Corrigan:1996nt} on the half infinite line.
In \cite{Mann:2006rh} the construction was generalized for the principal chiral model (PCM) on the finite line (open strings), and
was used to find integrable D-brane configurations for the bosonic sector of $\mathrm{AdS}_5\times \mathrm{S}^5$ background.
Some of the D-brane configurations, such as the maximal giant graviton, were shown to be integrable, while the method failed in other cases such as the $\mathrm{AdS}_4\times \mathrm{S}^2$ configuration.
The success of the construction is only a sufficient condition for integrability, thus a  failure of the construction does not imply that the model is not integrable.
While analyzing the bosonic sector, the fermionic sector was ignored.

In this paper we generalize the analysis of \cite{Mann:2006rh} in two directions. First, we consider the complete sectors including the fermionic degrees of freedom, i.e. we consider the GS$\s$M on supercoset backgrounds. Second, we do not limit ourselves to using reflection matrices, rather we consider more general boundary conditions involving all involutive automorphisms of the superalgebra.
We express the boundary conditions in terms of relations between the Maurer-Cartan one form's components with respect to the $\mathbb{Z}_4$ automorphism.
We find that in case where the boundary conditions break half of the supersymmetry, they can be written as
\be
a(z)=\O(\bar a(z^{-1})) \ ,
\ee
where $a(z)$ is the flat connection in the fixed frame, $z$ is the spectral parameter, $a=a_\tau+a_\s$, $\bar a=a_\tau-a_\s$ and $\O$ is an involutive metric preserving automorphism (not to be confused with the $\mathbb{Z}_4$ automorphism).
In these cases we can construct a generating function for an infinite set of non-local conserved charges (similar to the methods described above).

We use the method to find integrable D-brane configurations for $\mathrm{AdS}_5\times \mathrm{S}^5$ and $\mathrm{AdS}_4\times \mathbb{C}\mathrm{P}^3$ backgrounds.
In the case of $\mathrm{AdS}_5\times \mathrm{S}^5$ we find that the configurations that were shown to be integrable in \cite{Mann:2006rh} (i.e. the maximal giant graviton and $\mathrm{AdS}_5\times \mathrm{S}^3$) remain integrable when considering also the fermionic degrees of freedom.
On the gauge theory side, configurations of open strings ending on giant gravitons are known to have integrable structure \cite{Berenstein:2005vf}\cite{Berenstein:2005fa}\cite{Berenstein:2006qk}\cite{Hofman:2007xp}\footnote{See also \cite{Ahn:2010xa}\cite{MacKay:2010ey}\cite{MacKay:2010zb}\cite{Palla:2011eu} for some recent developments and \cite{Zoubos:2010kh} for a recent review on open boundaries and references therein.}.
In addition, we show that the $\mathrm{AdS}_4\times \mathrm{S}^2$ configuration is integrable by using outer automorphism of the PSU$(2,2|4)$
superalgebra.
On the gauge theory side, an evidence for integrability of the $\mathrm{AdS}_4\times \mathrm{S}^2$ configuration at weak coupling was presented in \cite{DeWolfe:2004zt}. However, attempters to prove integrability at strong coupling were not successful \cite{Mann:2006rh}. In \cite{Correa:2008av} it was suggested that the integrable structure found at one-loop in \cite{DeWolfe:2004zt} is accidental.
Recently, it was realized in \cite{Correa:2011nz}\cite{MacKay:2011zs}\cite{Regelskis:2011fa} that this configuration is indeed integrable using achiral boundary conditions in the scattering theory and achiral automorphism for the PCM.
We will find rather general solutions for automorphisms and construct more examples of integrable D-brane configurations.

The paper is organized as follows.
In section \ref{sec:Boundary conditions for Green-Schwarz sigma-models} we construct a class of boundary conditions for the GS$\s$M.
We check the consistency of the boundary conditions with by analyzing kappa symmetry and supersymmetry.
In section \ref{sec:Integrable D-brane configurations} we construct the generating function for the conserved charges, and show that the existence of such a function is equivalent to the boundary conditions found in section \ref{sec:Boundary conditions for Green-Schwarz sigma-models} if half of the supersymmetry is conserved.
In section \ref{sec:Integrable configurations for AdS5xS5} we use the construction for the $\mathrm{AdS}_5\times \mathrm{S}^5$ background and in section \ref{sec:Integrable configurations for the AdS4xCP3} for the $\mathrm{AdS}_4\times \mathbb{C}\mathrm{P}^3$ background.
Section \ref{sec:Discussion} is devoted to a discussion and outlook.
Details of the PSU$(2,2|4)$ and OSP$(6|4)$ superalgebras are given in appendices \ref{ap:PSU224SCA} and \ref{ap:OSP64SCA}, respectively.
In appendix \ref{ap:more_parameterizations} we provide a parameterizations of the $\mathrm{AdS}_5\times \mathrm{S}^5$ background.

\section{Boundary conditions for Green-Schwarz sigma-models}\label{sec:Boundary conditions for Green-Schwarz sigma-models}
The GS$\s$M action on a semi-symmetric spaces $G/H$ is given by \cite{Metsaev:1998it}
\be
S=\int d^2\s \Str(J^{(2)}\wedge \ast J^{(2)} + J^{(1)}\wedge J^{(3)}) \ ,
\ee
where $J=g^{-1}dg, g\in G$ is the Maurer-Cartan one-form,  and
$$
J^{(a)}=\frac{1}{4}(1+(-i)^a\hat\O+((-i)^a\hat\O)^2+((-i)^a\hat\O)^3)J \ ,
$$
with $a=0,..,3$ and $\hat\O$ is the $\mathbb{Z}_4$ automorphism map.

Next we derive the EOM's and find the boundary term. We note that
\be
\d J=\d(g^{-1}dg)=-g^{-1}\d g J+g^{-1}d\d g=[J,\D]-\D d(\cdot)+d(\D \cdot) \ ,
\ee
where we defined $\D\equiv g^{-1}\d g$. Therefore,the variation of the action is
\be
\d S
=\int \Str\bigg(\d J\wedge (2\ast J^{(2)}+J^{(3)}-J^{(1)})\bigg)
\ee
$$
=\int \Str\bigg(-\D(J\wedge (2\ast J^{(2)}+J^{(3)}-J^{(1)})+(2\ast J^{(2)}+J^{(3)}-J^{(1)})\wedge J
+d(2\ast J^{(2)}+J^{(3)}-J^{(1)}))
$$
$$
+d(\D(2\ast J^{(2)}+J^{(3)}-J^{(1)}))\bigg) \ .
$$
The equations of motion read
\be\label{eq:GSEOM}
d\ast J^{(2)}
+J^{(0)}\wedge\ast  J^{(2)}
+\ast J^{(2)}\wedge  J^{(0)}
+J^{(3)}\wedge J^{(3)}
-J^{(1)}\wedge J^{(1)}
=0 \ ,
\ee
$$
J^{(2)}\wedge (\ast J^{(3)}-J^{(3)})
+(\ast J^{(3)}-J^{(3)})\wedge J^{(2)}
=0 \ ,
$$
$$
J^{(2)}\wedge (\ast J^{(1)}+J^{(1)})
+(\ast J^{(1)}+J^{(1)})\wedge J^{(2)}
=0 \ ,
$$
where we made used the MC equations, $d J+J\wedge J=0$. The boundary term is
\be
\d S_{B}=\int \Str\bigg(d(\D(2\ast J^{(2)}+J^{(3)}-J^{(1)}))\bigg)
\ee
$$
~~~~~~~~~~~~~=\int d\tau \Str\bigg(\D(-2\sqrt{-h}J^{(2)}_\s-J^{(3)}_\tau+J^{(1)}_\tau)\bigg)\bigg|^{\s=\pi}_{\s=0} \ ,
$$
where $h$ is the determinant of the induced metric $h_{\a\b}(\tau,\s)$.
Let us define $J_\d \equiv \D = g^{-1}\d g$, that we find in the conformal gauge
\be\label{eq:BoundaryTerm}
\d S_{B}=\int d\tau \Str\bigg(-2J^{(2)}_\d J^{(2)}_\s-J^{(1)}_\d J^{(3)}_\tau+J^{(3)}_\d J^{(1)}_\tau\bigg)\bigg|^{\s=\pi}_{\s=0}.
\ee

Next, we express boundary conditions for which the boundary term vanishes in terms of the currents.
We use conformal coordinates, $\mathrm{z}=\tau+\s,\mathrm{\bar z}=\tau-\s$ so that $J\equiv J_\tau+J_\s$ and $\bar J\equiv J_\tau-J_\s$.
The boundary term (\ref{eq:BoundaryTerm}) vanishes for the following boundary conditions\footnote{Gluing conditions in CFT's in terms of the currents were introduced in \cite{Ishibashi:1988kg}. There are many cases where the gluing conditions were given in terms of automorphisms in the analysis of D-branes, e.g. for Gepner-models \cite{Recknagel:1997sb}, WZW-models on group manifolds \cite{Alekseev:1998mc} and supergroup manifolds \cite{Creutzig:2007jy},\cite{Creutzig:2008ag}, PCM's on symmetric spaces \cite{MacKay:2001bh} etc.}:
\be\label{eq:boudaryconditions}
J^{(2)}=\O(\bar J^{(2)}),\quad
J^{(1)}=\O(\bar J^{(3)}),\quad
J^{(3)}=\O(\bar J^{(1)}),\quad
\mathrm{at}\quad
\s=0,\pi
\ee
with $\O$ a metric preserving involutive automorphism map\footnote{Note that the automorphism map $\O$ is not related to the $\mathbb{Z}_4$ automorphism which we denote by $\hat\O$.} ($\Str(\O(A) \O(B))=\Str(A B)$, $\O^2=1$).
These boundary conditions leave the action invariant.
Equation (\ref{eq:boudaryconditions}) may be summarized using the moving frame's flat-connection $A(z)$ as
\be\label{eq:MovingFCTrans}
A(z)-J^{(0)}=\O(\bar A(z^{-1})-\bar J^{(0)}) \quad
\mathrm{at}\quad
\s=0,\pi
\ee
with\footnote{Taking $A(z)$ is related to $\bar A(z^{-1})$ through a parity transformation \cite{arutyunov-2009}.}
\cite{Bena:2003wd}\cite{Beisert:2005bm}
\be\label{eq:flatconnectionZ}
A(z)-J^{(0)}=z J^{(1)}+z^{-2} J^{(2)}+z^{-1} J^{(3)},
\ee
$$
\bar A(z)-\bar J^{(0)}=z \bar J^{(1)}+z^{2} \bar J^{(2)}+z^{-1} \bar J^{(3)}.
$$
$z\in \mathbb{C}$ the spectral parameter (not to be confused with the conformal coordinate $\mathrm{z}$).

In order for this boundary condition to be consistent it should respect the kappa-symmetry of the action.
Kappa symmetry acts as $g\rightarrow g e^{\hat\k}$ \cite{Metsaev:1998it}\cite{Adam:2007ws} with
\be\label{eq:kappa}
\hat\k=[J^{(2)}_\a,\k_{(1)}^\a]+[J^{(2)}_\a,\k_{(3)}^\a]
\ee
so the kappa-symmetry variation is $\D=g^{-1}\d g=\hat\k$. If $\O(\hat\k)=\hat\k$, the boundary term vanishes.
Let us act on (\ref{eq:kappa}) with $\O$, then
\be
\O(\hat\k)= \O([J^{(2)}_\a,\k_{(1)-}^\a]+[J^{(2)}_\a,\k_{(3)+}^\a])
\ee
$$
=[J^{(2)}_\tau, \O(\k_{(1)-}^\tau)]
-[J^{(2)}_\s, \O(\k_{(1)-}^\s)]
+[J^{(2)}_\tau, \O(\k_{(3)+}^\tau)]
-[J^{(2)}_\s, \O(\k_{(3)+}^\s)],
$$
where $\k_{\pm}^\a=P_{\pm}^{\a\b}\k_\b=\frac{1}{2}(h^{\a\b}\pm \e^{\a\b}/\sqrt{-h})\k_\b$.
Requiring $\O(\hat\k)=\hat\k$ we get
\be
\O(\k)=\bar \k,\quad
\mathrm{with}\quad \O^2=1,\quad
\mathrm{and}\quad
\k=\k_{(1)-}+\k_{(3)+}.
\ee
Thus, we conclude that our boundary conditions (\ref{eq:boudaryconditions}) are consistent\footnote{We assumed that the automorphism acts as a linear transformation, see \cite{1985IzMat..24..539S} for a classification of superalgebras automorphisms.}.

Let us now consider the supersymmetry breaking.
The supersymmetry transformation is given by $g\rightarrow e^\e g$, $\e=\e^1+\e^3\in \mathfrak{g}_{\bar 1}$, so $g^{-1}\d g= g^{-1}\e g$.
On the boundary we should have $\O(g^{-1}\d g)=g^{-1}\d g$.
We will encounter two types of automorphisms \cite{1985IzMat..24..539S}, one in which $\O(A B)=\O(A)\O(B)$ and the second in which $\O(A B)=-\O(B)\O(A)$. In the first case, if $\O(g)=g$ we should have $\O(\e)=\e$ which breaks half of the supersymmetry, since the relation is linear. In the second case, if $\O(g)=-g^{-1}$ we also find that $\O(\e)=\e$ which breaks half of the supersymmetry.

If the configuration breaks half of the supersymmetry as explained above, then equation (\ref{eq:boudaryconditions}) may be rewritten in the form
\be\label{eq:BC}
a(z)=\O(\bar a(z^{-1})) \quad
\mathrm{at}\quad
\s=0,\pi
\ee
with
\be\label{eq:flatconnectionZ}
a(z)=(z-1)j^{(1)}+(z^{-1}-1)j^{(3)}+(z^{-2}-1)j^{(2)},
\ee
$$
\bar a(z)=(z-1)\bar j^{(1)}+(z^{-1}-1)\bar j^{(3)}+(z^{2}-1)\bar j^{(2)},
$$
the flat connection in the fixed-frame and $j^{(m)}=g J^{(m)} g^{-1}$.

Finally, we note that the boundary conditions do not modify the Virasoro constraint
\be
\Str(J^{(2)}_\a J^{(2)}_\b-\frac{1}{2}g_{\a\b}g^{\g\d}J^{(2)}_\g J^{(2)}_\d)=0
\ee
on the boundary.

\section{Integrable D-brane configurations }\label{sec:Integrable D-brane configurations}
Adding boundaries to $(1+1)$-dimensional integrable models (originally with periodic boundary conditions) may break the integrable structure, since the monodromy matrix ceases to be conserved.
It is sometimes possible to construct a generating function for an infinite set of conserved charges for models with boundaries
by using the bulk transition matrix together with appropriate reflection matrices \cite{Sklyanin:1987bi}\cite{Sklyanin:1988yz}.
The construction was used for O$(N)$ sigma models on the half infinite line in \cite{Corrigan:1996nt}, and was generalized to the finite line (open strings) in \cite{Mann:2006rh} for the bosonic principal chiral model (PCM).
In this section we generalize the procedure for Green-Schwarz sigma-models (GS$\s$M's) on supercoset backgrounds.

The GS$\s$M \cite{Metsaev:1998it} has a flat-connection one-form $a(z)$, namely $d a(z)+a(z)\wedge a(z)=0$ with $z\in \mathbb{C}$ the spectral parameter \cite{Bena:2003wd}. This property allows one to construct the transition matrix given by \cite{Faddeev:1987ph}
\be
T(\s_2,\s_1;z)=\mathcal{P} e^{\int_{\s_1}^{\s_2} d\s a_{\s}(\s;z)} \ ,
\ee
where $\mathcal{P}$ is the path ordering symbol.
Using the flatness of $a(z)$, it follows that the transition matrix satisfies
\be\label{eq:TransitionDer}
\p_\tau T(\s_2,\s_1;z)=a_{\tau}(\s_2;z)T(\s_2,\s_1;z)-T(\s_2,\s_1;z)a_{\tau}(\tau_1;z).
\ee
Defining the monodromy matrix $T_\g(z)\equiv T(2\pi,0;z)$, we see that if the boundary conditions are periodic (namely, $a(0;z)=a(2\pi;z)$) then
\be
\p_\tau \Str(T_\g(z))=0,
\ee
so $\Str(T_\g(z))$ is a generating function for integrals of motion. Note, that in fact these boundary conditions imply $\Str(T_\g(z)^n)=0$ for any $n\in\mathbb{Z}_+$.
Obviously, without the periodic boundary conditions the monodromy matrix is generally not conserved. Next we describe how to construct a generating function for integrals of motion in the case of open string boundary conditions.

We start by following the construction given in \cite{Mann:2006rh} (though we use different notations).
First we define the function
\be\label{eq:GeneratingFunc}
T(z)\equiv U_0T^{-1}(\pi,0;z^{-1}) U_\pi T(\pi,0;z).
\ee
The $U_{0/\pi}$ matrices are the reflection matrices at the two ends of the string. We take them to be constant matrices as explained in \cite{Mann:2006rh}. Generally $U$ may depend on $z$, but as we explain at the end of the section, the relevant $U$'s for our constructions are $z$-independent.
In \cite{Mann:2006rh} our $T^{-1}(\pi,0;z^{-1})$ is denoted by a function $T_R$, which is equivalent to our $T^{-1}$ at the bosonic level.
Furthermore, note that we work with a $\mathbb{Z}_4$ coset sigma-model while in \cite{Mann:2006rh} the PCM was analyzed, so the flat-connection takes a different form, and also our string range is $\s\in(0,\pi)$.

Next we show under what conditions $\Str(T(z))$ is conserved.
Using (\ref{eq:TransitionDer}) we differentiate (\ref{eq:GeneratingFunc}) with respect to $\tau$, so
\be
\p_\tau T(z)=
U_0 T^{-1}(\pi,0;z^{-1}) U_\pi \bigg(a_\tau(\pi;z)T(\pi,0;z)-T(\pi,0;z)a_\tau(0;z)\bigg)
\ee
$$
~~~~~~~~~~~~~~~+U_0\bigg(a_\tau(0;z^{-1})T^{-1}(\pi,0;z^{-1})-T^{-1}(\pi,0;z^{-1})a_\tau(\pi;z^{-1}) \bigg)U_\pi T(\pi,0;z).
$$
Requiring $\p_\tau\Str(T(z))=0$ we find that
\be\label{eq:IntCond}
 U_\pi a_\tau(\pi;z)-a_\tau(\pi;z^{-1}) U_\pi =0,\quad
 U_0 a_\tau(0;z^{-1})-a_\tau(0;z) U_0 =0.
\ee
As in the closed string, the boundary conditions (\ref{eq:IntCond}) imply that $\p_\tau\Str(T(z)^n)=0$, where $n$ is a positive integer.
If we identify
\be
\O_U(x)\equiv U x U^{-1},\quad
x\in \mathfrak{g}
\ee
as an automorphism, then (\ref{eq:IntCond}) defines a consistent boundary condition given that  $\O^2=1$ and $\O_U(g)=g$.
To be more precise, (\ref{eq:BC}) with $\O_U$ should also be satisfied so that the action will be invariant. Throughout the analysis we assume that the reflection matrices $U$ are invertible.
To summarize, we find that if
\be\label{eq:intCondRefMat}
\O_{U_0}(a_\tau(0,z))=a_\tau(0,z^{-1}),\quad
\mathrm{and}
\quad
\O_{U_\pi}(a_\tau(\pi,z^{-1}))=a_\tau(\pi,z),
\ee
the model is classically integrable.
Plugging the GS$\s$M's flat-connection
\be\label{eq:flatconnectionZinv}
a(z)=(z-1)j^{(1)}+\frac{1}{2}(z-z^{-1})^2 j^{(2)}+(z^{-1}-1)j^{(3)}-\frac{1}{2}(z^2-z^{-2}) \ast j^{(2)},
\ee
where $j^{(m)} =g J^{(m)} g^{-1}$, in (\ref{eq:intCondRefMat}),
we get the explicit relations for integrability
\be\label{eq:integrabilitycondition1}
[ U,j^{(2)}_\tau]=\{ U,j^{(2)}_\s\}=0,
\ee
\be\label{eq:integrabilitycondition2}
j^{(1)}_\tau= U^{-1}j^{(3)}_\tau U,\quad
j^{(3)}_\tau= U^{-1}j^{(1)}_\tau U,
\ee
\be\label{eq:integrabilitycondition3}
j^{(1)}_\s= -U^{-1}j^{(3)}_\s U,\quad
j^{(3)}_\s= -U^{-1}j^{(1)}_\s U,
\ee
with $U^2=\pm 1$. The last equation follows from the invariance of the action, or equivalently, the EOM. If $\O_{U}(g)=g$, then the same equations apply for $J$ (the moving frame current).

The result (\ref{eq:intCondRefMat}) calls for a more general condition for integrability, that is, we may consider a general automorphisms not necessarily restricting ourselves to matrix conjugation\footnote{Automorphisms of superalgebras are classified in \cite{1985IzMat..24..539S}}.
This is what we will do next.
We define the matrices
\be
T(\pi,0;z)=\mathcal{P}\exp\left(\int_0^\pi a(z)\right),\quad
T_{\tilde\O}(0,\pi;z^{-1})=\mathcal{P}\exp\left(\int_0^\pi {\tilde\O}(a(z^{-1}))\right),
\ee
with ${\tilde\O}$ an automorphism map, and define a new object
\be
T(z)\equiv U_0 T^{-1}_{\tilde\O}(\pi,0;z^{-1}) U_\pi T(\pi,0;z)
\ee
(so taking $\tilde \O$ to be the identity map, we return to the old definition (\ref{eq:GeneratingFunc})).
Differentiating $\Str(T(z))$ as before, we end up with the integrability conditions
\be
 U_\pi a_\tau(\pi;z)-\tilde\O(a_\tau(\pi;z^{-1})) U_\pi =0,\quad
 U_0 \tilde\O(a_\tau(0;z^{-1}))-a_\tau(0;z) U_0 =0.
\ee
From now on we will consider only the $\s=0$ boundary conditions since the $\s=\pi$ boundary conditions are equivalent.
Let us also introduce the notation $\O (x)\equiv U_0 \tilde\O(x) U_0^{-1}$, which is a composition of two automorphisms.
Finally, combining the integrability conditions together with consistency of the equations of motion we obtain the conditions
\be
a_\tau(0;z)=\O(a_\tau(0;z^{-1})),\quad
a_\s(0;z)=-\O(a_\s(0;z^{-1}))\quad
\ee
In order for these boundary conditions to be consistent, it is obvious that $\O$ should be an involutive automorphism.
This definition is not restricted to similarity transformation automorphisms as the previous one.
In the next section we will show that allowing a general automorphism, changes significantly the classification of integrable boundary conditions.

Comparing to the boundary conditions we found for half-BPS D-branes, we conclude that all half-BPS D-branes described by the gluing conditions, given in the previous section, are also integrable.

Let us comment about the $z$-independence of $U$.
The explicit relations (\ref{eq:integrabilitycondition1}-\ref{eq:integrabilitycondition3}) follows from (\ref{eq:intCondRefMat}) by identifying the $z$-dependent coefficients of the flat-connection. This means that unless we have the same coefficient functions in front of the $\mathbb{Z}_4$ graded one-forms, they will have to vanish by the boundary conditions (both their $\s$ and $\tau$ components).
If $U$ depends on $z$, it will change the coefficients according to their charge under $U$, and we do not find such $U$ which will transform all the coefficient functions by just interchanging between them up to some constants, in a way that is consistent with (\ref{eq:BC}).

\section{Integrable configurations for the $\mathrm{AdS}_5\times \mathrm{S}^5$ Background}\label{sec:Integrable configurations for AdS5xS5}
In this section we will first present some general results for the $\mathrm{AdS}_5\times \mathrm{S}^5$ background in global coordinates.
We will then take a closer look at the giant-graviton configuration in order to demonstrate how the prescription for finding the integrable structure works.
Next we will consider the Karch-Randall configuration in some detail, since we will use there a different type of automorphism involving the supertranspose operation, which allows us to prove integrability of the configuration (some previous attempt to find integrable structure using reflection matrices failed \cite{Mann:2006rh}\cite{Correa:2008av}, although recent attempts on the gauge theory side \cite{Correa:2011nz} were successful, see also \cite{MacKay:2011zs}\cite{Regelskis:2011fa}).
We will outline general solutions for allowed automorphisms and integrable D-brane configurations in $\mathrm{AdS}_5\times \mathrm{S}^5$.
The required details of the PSU$(2,2|4)$ superconformal algebra are given in appendix \ref{ap:PSU224SCA}.

\subsection{The $\mathrm{AdS}_5\times \mathrm{S}^5$ bosonic background in global coordinates }
The $\mathrm{AdS}_5\times \mathrm{S}^5$ background metric in global coordinates is given by $ds^2=ds^2_{\mathrm{AdS}_5}+ds^2_{\mathrm{S}^5}$ with
\be
ds^2_{\mathrm{AdS}_5}=d\r^2
-\cosh^2\r dt^2
+\sinh^2\r (d\a^2
+\sin^2\a d\b^2
+\cos^2\a d\g^2),
\ee
\be
ds^2_{\mathrm{S}^5}=d\t^2
+\sin\t ^2 d\phi^2
+\cos^2\t (d\psi^2
+\sin^2\psi d\eta^2
+\cos^2\psi d\varphi^2).
\ee
Concentrating on the bosonic sector, we have only the kinetic term in the action, $\mathrm{Str}(J^{(2)}\wedge \ast J^{(2)})$, where the $\mathcal{H}_2$ space is spanned by the $P$'s (translation generators).
In order to get the desired metric we use the coset element representative $g=g_{\mathrm{AdS}_5}g_{\mathrm{S}^5}$ with
\be
g_{\mathrm{AdS}_5}=e^{-P_{0}t}e^{-J_{13}\g}e^{J_{24}\b}e^{-J_{14}\a}e^{P_{1}\r},\quad
\ee
\be
g_{\mathrm{S}^5}=e^{-J_{79}\phi}e^{P_{8}\varphi}e^{J_{56}\eta}e^{P_{6}\psi}e^{P_{7}\t}.
\ee
The bosonic sector's $J^{(2)}$ is given by
\be\label{eq:J2BOSINIC}
J^{(2)}=
P_0\cosh\r dt
+P_1d\r
+P_2\sinh\r\sin\a d\b
+P_3\sinh\r\cos\a d\g
+P_4\sinh\r d\a
\ee
$$
+P_5\cos\t\sin\psi d\eta
+P_6\cos\t d\psi
+P_7d\t
+P_8\cos\t\cos\psi d\varphi
+P_{9}\sin\t d\phi.
$$
The vielbeins are given by
\be\label{eq:vielbein}
e^0_t=\cosh\r,\quad
e^1_\r=1,\quad
e^2_\b=\sinh\r\sin\a,\quad
e^3_\g=\sinh\r\cos\a,\quad
e^4_\a=\sinh\r,\quad
\ee
$$
e^5_\eta=\cos\t\sin\psi,\quad
e^6_\psi=\cos\t,\quad
e^7_\t=1,\quad
e^8_\varphi=\cos\t\cos\psi,\quad
e^{9}_\phi=\sin\t.
$$

We constructed the $P_{\hat a}$ and $J_{\hat a\hat b}$ matrices using the $4\times 4$ gamma matrices (given in (\ref{eq:smallgammas})) and their commutators respectively, so we will be able to use their Clifford algebra when computing their anti-commutation relations.

\subsection{The Maximal Giant Graviton D3-Brane}
The maximal giant graviton \cite{McGreevy:2000cw} is defined by the boundary conditions
\be
\r=0,\quad
\theta=0,\quad
t=\phi=\phi(\tau),
\ee
and Neumann boundary conditions for the rest of the coordinates, so the D-brane geometry is $\mathbb{R}^1\times \mathrm{S}^3$.
At the boundary, the bosonic sector's current (\ref{eq:J2BOSINIC}) is reduced to
\be
J^{(2)}=
P_0 dt
+P_1 d\r
+P_5\sin\psi d\eta
+P_6d\psi
+P_7d\t
+P_8\cos\psi d\varphi,
\ee
and in worldsheet components to
\be
J^{(2)}_\tau=
P_0 \p_\tau t
+P_5\sin\psi \p_\tau \eta
+P_6\p_\tau \psi
+P_8\cos\psi \p_\tau \varphi,
\ee
\be
J^{(2)}_\s=
P_1\p_\s \r
+P_7\p_\s \t.
\ee
In order to satisfy the integrability conditions we should take $ U =a P_0 + b J_{7,9}$, where $a,b\in \mathbb{C}$ are arbitrary for the moment (this result follows easily by using the commutation relations together with the Clifford algebra).
On the boundary $[U,g]=0$, so the lower-case $j^{(2)}=g J^{(2)} g^{-1}$ also satisfy
\be
[j^{(2)}_\tau, U ]=\{j^{(2)}_\s, U \}=0.
\ee

Next, we introduce the fermions.
We take the super-coset representative to be
\be
g=g_F g_B,\quad
g_B=g_{\mathrm{AdS}_5}g_{\mathrm{S}^5},\quad
g_F=e^F,
\ee
where
\be
F=\t\cdot Q
=\t_{I \a \a' a}Q_{I\b\b' b} C^{\a\b}C'^{\a'\b'}(i\s_2)^{ab}S^{IJ}
\ee
and
\be
\t_{I \a \a' 1}=\t_{I \a \a'},\quad
\t_{I \a \a' 2}=0,\quad
Q_{I\b\b' 1}=0,\quad
Q_{I\b\b' 2}=-Q_{I\b\b'}.
\ee
$\a,\a',a$ and $I$ are the AdS$_5$-spinor, S$^5$-spinor, chirality and internal indices respectively. The $32\times 32$ gamma matrices that act on the $\a,\a',a$ indices are tensor products of the $\a,\a'$ and $a$ spaces (see appendix \ref{ap:PSU224SCA}).
$\t$ and $Q$ have opposite chirality, while $\t_1$ and $\t_2$ have the same chirality, and so do the $Q$'s.
We take $S^{IJ}$ to be diagonal $2\times 2$ matrix.

The current $J$ decomposes to
\be
J=g^{-1}_B (g^{-1}_F d g_F) g_B+g^{-1}_B d g_B \ ,
\ee
where we have already analyzed above the second term, and realized that $g_B$ on the boundary commutes with $U$.
The $j^{(1)}_\tau$ and $j^{(3)}_\tau$ should satisfy
\be
j^{(1)}_\tau= U  j^{(3)}_\tau  U ^{-1}, \quad
[U^2,j^{(1)}_\tau]=[U^2,j^{(3)}_\tau]=0.
\ee
In order for these relations to be satisfied, they should be satisfied for the moving frame's currents, namely
\be\label{eq:JF_Int_cond}
J^{(1)}_\tau= U  J^{(3)}_\tau  U ^{-1}, \quad
[U^2,J^{(1)}_\tau]=[U^2,J^{(3)}_\tau]=0.
\ee

We note that the matrix transformation given by $U$, induces a transformation in the spinor space\footnote{We use the $\e_{IJ}$ symbol since $Q^1$ and $Q^2$ should transform we a relative sign. This can be seen we noting that $Q^1_{\a\a'}=i\S Q^2_{\a\a'}$ where $\S=\mathrm{diag}(+,+,+,+,-,-,-,-)$, and $[U,\S]=0$.}
\be
U Q_{I\a\a' a} U^{-1}=U^{(s)}_{\a\a' a}{}^{\b\b' b}\e_{IJ} Q_{J\b\b' b}=(\g_A)_\a{}^\b (\g_S)_{\a'}{}^{\b'}\s_a{}^b\e_{IJ} Q_{J\b\b' b}.
\ee
We require $F$ to be invariant under the similarity transformation ($U F U^{-1} = F$), so the $\t$'s should satisfy on the boundary
\be\label{eq:FermionsBC}
\t^{\b\b' b}_J=\t^{\a\a' a}_I U^{(s)}_{\a\a' a}{}^{\b\b' b}\e_{IJ}
=-\t^{\g\g' c}_J U^{(s)}_{\g\g' c}{}^{\a\a' a} U^{(s)}_{\a\a' a}{}^{\b\b' b} \ .
\ee
In order not to reduce further the number of degrees of freedom we should have $(U^{(s)})^2=-1$, with
$U^{(s)}=\g_A\otimes \g_S\otimes \s$.

The invariance of $F$ on the boundary guarantees that $[U,g_F]=0$ .
The boundary condition for $\p_\tau\t$ is the same as (\ref{eq:FermionsBC}), so the first condition in (\ref{eq:JF_Int_cond}) is satisfied.
Taking the boundary condition for $\p_\s\t$ to be the same as (\ref{eq:FermionsBC}) but with a minus sign on the LHS, then $UJ^{(1)}_\s U^{-1}=-J^{(3)}_\s$ is satisfied.
Finally, the second part of (\ref{eq:JF_Int_cond}) is satisfied if $U_A^2=U_S^2\propto 1_{4\times 4}$.

Since we chose the bosons and fermions boundary conditions such that
$$U\p_\tau F U^{-1}=\p_\tau F,\quad U\p_\s F U^{-1}=-\p_\s F,\quad U F U^{-1}=F,\quad U g_B U^{-1}=g_B,$$
it follows that the fermions contribution to $J^{(2)}$ does not modify (\ref{eq:integrabilitycondition1}).

These conditions fix the $a$ and $b$ coefficients up to normalization and relative sign to be
\be
U=2 P_0-i 2^3 P_5 P_6 P_8,
\ee
in the superalgebra basis, and
\be
U^{(s)}\otimes \e=\G^0\G^5\G^6\G^8\otimes \e=\g^0\otimes \g^5\g^6\g^8\otimes (i\s^3)\otimes \e,
\ee
in the spinor basis, so the chirality is preserved, and $U^2=-1_{8\times 8}$, so all the conditions for integrability are satisfied and consistent.
We can identify the gamma-matrices indices with the Neumann directions using the vielbeins given in (\ref{eq:vielbein}), so $0,5,6,8$ are associated with the $t,\eta,\psi,\varphi$ directions respectively.
The Majorana condition is also satisfied upon the identification (\ref{eq:FermionsBC}) will be explained later.

\subsection{Karch-Randall D5-Brane}
Consider the Karch-Randall D5-Brane \cite{Karch:2000gx} wrapping $\mathrm{AdS}_4\times \mathrm{S}^2$. In the analysis of \cite{Mann:2006rh} it is shown that applying the finite line integrability procedure fails for this configuration, but recent attempts have shown that it is indeed integrable upon using \textit{twisted} or \textit{achiral} boundary conditions \cite{Correa:2011nz}\cite{MacKay:2011zs}\cite{Regelskis:2011fa}. We will show that this configuration is integrable by using an automorphism of the form $\O(x)=-U x^{st} U^{-1}$.

 We use the parametrization given in appendix \ref{ap:more_parameterizations} for the Poincar\'{e} and spherical coordinates, and take the boundary conditions
\be
x_2=0,\quad
\t_7=\t_8=\t_{9}=0.
\ee
The bosonic sector's current $J^{(2)}$ in worldsheet coordinates is reduced to
\be
J^{(2)}_\tau=\frac{1}{y}(P_0 \p_\tau t+P_1 \p_\tau x_1+P_3 \p_\tau x_3-P_4 \p_\tau y)
+(P_{8}\p_\tau \t_{8}+\cos\t_{8}(P_{5}\p_\tau \t_{5}))
\ee
\be
J^{(2)}_\s=P_2 \frac{\p_\s x_2}{y}
+P_{9}\p_\s \t_{9}+P_{6}\p_\s \t_{6}+P_{7}\p_\s \t_{7}.
\ee
The automorphism which respects the boundary conditions is
\be
\O(x)=-U x^{st} U^{-1},\quad
U=2 P_4 - 4 P_6 P_7.
\ee
This automorphism transformation can be checked to be involutive, and satisfying $\O(A B)=-\O(B)\O(A)$ and $\O(g_B)=-g_B^{-1}$.
The spinors transformation is given by
\be
U^{(s)}=\g_5\g_6\otimes \g_0\g_1\g_3\g_4\otimes 1 \ ,
\ee
and we see that the supertranspose operation interchanges the role of the two spinor indices, $\a\leftrightarrow \a'$.
In any case, these gamma indices correspond to the Neumann directions, as can be read from the expressions for $J^{(2)}$ in appendix \ref{ap:PSU224SCA}.

\subsection{Other integrable configurations}
Any metric preserving involutive automorphism of the PSU$(2,2|4)$ superalgebra may serve in principle for the gluing of the currents on the boundary. The automorphisms for simple superalgebras were classified in \cite{1985IzMat..24..539S}. We consider two types of automorphisms,
\be
\O(x)=U x U^{-1},\quad
\O_{st}(x)=-U x^{st} U^{-1},\quad
U\in \mathrm{GL}(4|4)_{\bar 0}.
\ee
Let us consider $\O$ first. In this case we represent the $U$ matrix as a product of $P_a$ matrices plus a product of $P_{a'}$ matrices, that is $U=\prod_a c_a P_a + \prod_{a'} c_{a'}P_{a'}\equiv U_A+U_S$ with $c_a,c_{a'}\in \mathbb{C}$ for some set $\{a,a'\}$. The $P$ matrices represent the $4\times 4$ gamma matrices (\ref{eq:smallgammas}), so we can immediately find which $P$ commutes or anti-commutes with $U$, see table \ref{table:gammaU}.
Using the relation $\g^4\sim \g^{0}\g^{1}\g^{2}\g^{3}$, we can always use $U\sim~\g^{p+1,..,4}$ instead of $U\sim \g^{0,..,p}$. Then whenever $p$ is even, it is natural to consider $U\sim~\g^{p+1,..,4}$ since then the $a=P+1,..,4$ directions are the Neumann directions (the ones that commutes with $U$, so their translation symmetry is not broken). Then, we see that the number of Neumann directions is always odd for the AdS and sphere's subspace, and so totally even for the entire space as should be for IIB.

\begin{table}[ht]
\caption{Commutation and anti-commutation of the gamma matrices}
\centering
\scriptsize
\begin{tabular}{|l ||c| c| c| c| c|}
\hline\hline
{}&{}&{}&{}&{}&{} \\
$U$        & $\g^1$ & $\g^{12}$ & $\g^{123}$ & $\g^{1234}$ & $\g^{12340}$ \\
{}&{}&{}&{}&{}&{} \\ \hline
{}&{}&{}&{}&{}&{} \\
comm       & $\g^1$ & $\g^3,\g^4,\g^0$ & $\g^1,\g^2,\g^3$ & $\g^0$ & $\g^1,\g^2,\g^3,\g^4,\g^0$ \\
{}&{}&{}&{}&{}&{} \\ \hline
{}&{}&{}&{}&{}&{} \\
anti-comm  & $\g^2,\g^3,\g^4,\g^0$ & $\g^1,\g^2$ & $\g^4,\g^0$ & $\g^1,\g^2,\g^3,\g^4$ & $\emptyset$ \\
{}&{}&{}&{}&{}&{} \\
\hline
\end{tabular}
\begin{quote}
The first row represents the different $U$ matrices. The second (third) row gives the gamma matrices that (anti-)commutes with $U$.
The analysis is the same for the AdS and sphere's $4\times 4$ gamma matrices.
\end{quote}
\label{table:gammaU}
\end{table}
\normalsize

Next we find the allowed $U$ matrices. The consistency conditions are $U^{2}=\pm 1$, $\O$-should interchange generators
$\mathcal{H}_1\leftrightarrow \mathcal{H}_3 $, so should have $\hat\O(\O (X^1))=-i \O (X^1)$ (where $\hat\O( X^1)=i X^1$) and finally the Majorana condition should be preserved (since the total number of gamma matrices is even the chirality of the spinors is not changed). The second condition implies
\be
\hat\O(\O (X^1))=\hat\O(U X^1 U^{-1})
=i\hat\O(U^{-1})X^1 \hat\O(U)
=i\hat\O(U)X^1 \hat\O(U^{-1})
\ee
$$
=i K U^t K^{-1}X^1 K U^{t-1} K^{-1}
=i k U^t k^{-1} X^1 k U^{t-1} k^{-1}
$$
where we used $U^2=\pm 1$ in the third equality, and $\hat\O(x)=-K x^{st} K^{-1}$ (see appendix \ref{ap:PSU224SCA}).
We also defined
\be
K=\left(
    \begin{array}{cc}
      C & {} \\
      {} & i C \\
    \end{array}
  \right),\quad
k=\left(
    \begin{array}{cc}
      C & {} \\
      {} & C \\
    \end{array}
  \right),\quad
\S=\left(
    \begin{array}{cc}
      1 & {} \\
      {} & -1 \\
    \end{array}
  \right),
\ee
where $C$ is the charge conjugation matrix defined in appendix \ref{ap:PSU224SCA}.
Then it follows that $U$ must satisfy
\be\label{eq:condition2}
k U^t k^{-1}=\pm U\S.
\ee
Then we can use the charge conjugation matrix properties, namely $C\g^t_\m C^{-1}=\g_\m$ with $C^2=-1$, so
$C\g^t_{\m\n\r...} C^{-1}$ equals $\g_{\m\n\r...} $ for 1, 4 and 5 Lorentz indices and equals $-\g_{\m\n\r...} $ for 2 and 3 indices.
Let us denote $U=\tiny\left(
                   \begin{array}{cc}
                     U_A & 0 \\
                     0 & U_S \\
                   \end{array}
                 \right)
$,
so $U_A$ is a product of 1, 4 or 5 $\g$-matrices and $U_S$ a product of 2 or 3 $\g$-matrices, or vice-versa.
As explained above we trade a product of two and four $\g$-matrices with three and one $\g$-matrices respectively, so we are left with
(1,3), (3,1), (3,5) and (5,3) D-branes only (the notation gives the number of dimensions the D-brane fills in the AdS and sphere subspaces.

Next we consider the Majorana condition. The odd generators satisfy $Q^{I \dag}_{\a\a'}=k Q^I_{\a\a'} k^{-1}$, so if $Q^{1 \dag}_{\a\a'}=k Q^1_{\a\a'} k^{-1}$, then applying $\O$ we have
\be\O(Q^{1 \dag}_{\a\a'}) =\O(k) \O(Q^1_{\a\a'})\O( k^{-1}),
\ee
but $\O(Q^{1 \dag}_{\a\a'})$ is in $\mathcal{H}_3$ and should also satisfy
\be
\O(Q^{1 \dag}_{\a\a'}) =k \O(Q^1_{\a\a'})k^{-1},
\ee
so $\O(k)=\pm k$.
This relation can also be written as
\be
k U k^{-1}=\pm U,\quad
\mathrm{or}\quad
[k,U]=0\quad
\mathrm{or}\quad
\{k,U\}=0.
\ee
In practice it means that both $U_A$ and $U_S$ should commute or anti-commute with $C$.
Any $U_A$ ($U_S$) which contain in its product of $P$ generators $P_2$ or $P_4$ ($P_7$ or $P_9$) but not both anti-commutes with $C$, else it commutes with $C$.
Together with (\ref{eq:condition2}) it leaves us with
\be
U=\bigg\{\left(
  \begin{array}{cc}
    P_{a}/P_{a b c d e} & {} \\
    {} & P_{a'b'c'}/P_{a'\tilde b'\tilde c'} \\
  \end{array}
\right),
\left(
  \begin{array}{cc}
    P_{\tilde a} & {} \\
    {} & P_{a'b'\tilde c'} \\
  \end{array}
\right),
\ee
$$
~~~~~~~~\left(
  \begin{array}{cc}
    P_{a b c}/P_{a \tilde b \tilde c} & {} \\
    {} & P_{a'}/P_{a'b'c'\tilde d'\tilde e'} \\
  \end{array}
\right),
\left(
  \begin{array}{cc}
    P_{a b\tilde c} & {} \\
    {} & P_{\tilde a} \\
  \end{array}
\right)
\bigg\},
$$
where the notation is $P_{abc..}=P_a P_b P_c ...$.
The $U^2=\pm 1$ requirement fixes the relative coefficients $c_{\hat a}$ up to normalization, which leaves us with
\be
U=\bigg\{
i\left(
  \begin{array}{cc}
    P_{0}/P_{0 1 2 3 4} & {} \\
    {} & i P_{a'b'c'}/ i P_{a'\tilde b'\tilde c'} \\
  \end{array}
\right),
\left(
  \begin{array}{cc}
    P_{a} & {} \\
    {} & P_{a'b'c'}/ P_{a'\tilde b'\tilde c'} \\
  \end{array}
\right),
\left(
  \begin{array}{cc}
    P_{\tilde a} & {} \\
    {} & P_{a'b'\tilde c'} \\
  \end{array}
\right),
\ee
$$
\left(
  \begin{array}{cc}
    P_{0 b c}/P_{0 \tilde b \tilde c} & {} \\
    {} & i P_{a'}/ i P_{5 6 7 8 9} \\
  \end{array}
\right),
i\left(
  \begin{array}{cc}
    P_{a b c}/P_{a \tilde b \tilde c} & {} \\
    {} & P_{a'}/ P_{5 6 7 8 9} \\
  \end{array}
\right),
\left(
  \begin{array}{cc}
    P_{0 b\tilde c} & {} \\
    {} & i P_{\tilde a} \\
  \end{array}
\right),
i\left(
  \begin{array}{cc}
    P_{a b\tilde c} & {} \\
    {} & P_{\tilde a} \\
  \end{array}
\right)
\bigg\}
$$
We inserted a factor of $i$ in front of some of the matrices so that all of them are hermitian. The $P$'s should be understood to be normalized by a factor of 2 for each $P$, namely $P_{ab..}=(2P_a)(2P_b)...$.
This class of half-BPS integrable D-branes is consistent with the classification given in the literature e.g. \cite{Skenderis:2002vf}\cite{Sakaguchi:2003py}.

In table \ref{table:AdSnoSTDbranes} we give some of the possible configurations with $U$ satisfying the conditions described above.

\begin{table}[ht]
\caption{Integrable half-BPS D-brane configurations for $\mathrm{AdS}_5\times \mathrm{S}^5$ with $\O(x)=U x U^{-1}$.}
\centering
\scriptsize
\begin{tabular}{|l ||l| l|}
\hline\hline
$\mathbb{R}^{1,0}\times \mathrm{S}^3$             &$U=P_{0}+iP_{5,6,9}$         &$\r=0;\t=0$\\
$\mathbb{R}^{0,1}_+\times \mathrm{S}^3$           &$U=P_{1}+P_{5,6,9}$          &$t=0,\a=0;\g=0,\t=0$\\
$\mathrm{AdS}_3\times \mathrm{S}^1$  &$U=P_{0,1,4}+iP_{8}$          &$\g=\b=0;\t=\psi=0$\\
$\mathbb{H}^3\times \mathrm{S}^1$                 &$U=P_{1,2,3}+P_{8}$           &$t=0,\b=0;\t=\psi=0$\\
$\mathrm{AdS}_3\times \mathrm{S}^5$               &$U=P_{0,1,3}+iP_{5,6,7,8,9}$ &$\a=0$\\
$\mathbb{H}^3\times \mathrm{S}^5$                 &$U=P_{1,2,3}+P_{5,6,7,8,9}$  &$t=0,\b=0$\\
$\mathrm{AdS}_5\times \mathrm{S}^3$               &$U=P_{0,1,2,3,4}+iP_{5,6,8}$  &$\t=0$\\
\hline
\end{tabular}
\begin{quote}
We give some of the integrable D-brane configurations with the gluing conditions involving $\O(x)=U x U^{-1}$ automorphism.
The second column gives the $U$ matrices in terms of the superalgebra generators $P_{\hat a}$, the notation is $P_{a,b,c,...}=(2P_a)(2P_b)(2P_c)...$. For each boundary condition (\ref{eq:IntCond}) is satisfied with $\O(g)=g$ on the boundary.
\end{quote}
\label{table:AdSnoSTDbranes}
\end{table}
\normalsize

Next we consider the $\O_{st}$ automorphism.
First we note that in this case we should have $U=\pm \S U^t$ in order for the automorphism to be involutive.
Again $\O_{st}$ should interchange generators
$\mathcal{H}_1\leftrightarrow \mathcal{H}_3 $,
so we should have $\hat\O(\O_{st} (X^1))=-i \O_{st} (X^1)$ (where $\hat\O( X^1)=i X^1$).
Also the chirality and the Majorana condition should be preserved.
The second condition implies
\be
\hat\O(\O_{st} (X^1))
=-\hat\O(U^{-1})\hat\O(X^{1 st}) \hat\O(U)
=i\hat\O(U)X^{1 st} \hat\O(U^{-1})
\equiv i U X^{1 st} U^{-1}.
\ee
so we conclude that
\be
\hat\O(U)=\pm U^{-1},\quad
\ee
Similarly to the previous analysis, the Majorana condition requires
\be
k U k^{-1}=\pm U,\quad
\mathrm{or}\quad
[k,U]=0\quad
\mathrm{or}\quad
\{k,U\}=0.
\ee
Applying these conditions on automorphisms with $U=\prod_a c_a P_a + \prod_{a'} c_{a'}P_{a'}\equiv U_A+U_S$ as before, we find that the possible combinations are (0,2), (2,0), (2,4) and (4,2) branes. This result is also consistent with the classification of \cite{Skenderis:2002vf}\cite{Sakaguchi:2003py}\footnote{Note, however, that the D(-1)-brane is not in this class of integrable boundary conditions.} and preserve the chirality.
As for the previous type of automorphism, we can find several Half-BPS integrable D-branes, see table \ref{table:AdSSTDbranes} for examples.

\begin{table}[ht]
\caption{Integrable half-BPS D-brane configurations for $\mathrm{AdS}_5\times \mathrm{S}^5$ with $\O_{st}(x)=-U x^{st} U^{-1}$.}
\centering
\scriptsize
\begin{tabular}{|l ||l| l|}
\hline\hline
$\mathrm{AdS}_2 $                                 &$U=P_{3}+P_{7,9}$ &$\a=\b=\g=0;\t=\psi=\varphi=0$\\
$\mathrm{S}^2\subset\mathrm{S}^5$                 &$U=P_{2,4}+i P_{6}$ &$\r=0,t=0;\t_6=\t_7=\t_{9}=0$\\
$\mathbb{H}^2 $                                   &$U=P_{0}+iP_{7,9}$ &$t=0,\a=0;\t=\psi=\varphi=0$\\
$\mathrm{AdS}_2\times \mathrm{S}^4$               &$U=P_{0,2}+P_{7}$ &$x^i=0;\t_9=0$\\
$\mathbb{H}^2\times \mathrm{S}^4$                 &$U=P_{0}+iP_{6,8}$ &$t=0,\a=0;\t_{5}=0$\\
$\mathrm{AdS}_4\times \mathrm{S}^2$               &$U=P_{4}-P_{6,7}$ &$x^2=0;\t_{5}=\t_7=\t_8=0$\\
$\mathbb{H}^4\times \mathrm{S}^2$                 &$U=P_{1,3}+i P_{6}$ &$t=0;\t_6=\t_7=\t_{9}=0$\\
\hline
\end{tabular}
\begin{quote}
We give some of the integrable D-brane configurations with the gluing conditions involving $\O(x)=-U x^{st} U^{-1}$ automorphism.
The second column gives the $U$ matrices in terms of the superalgebra generators $P_{\hat a}$, the notation is $P_{a,b,c,...}=(2P_a)(2P_b)(2P_c)...$. For each boundary condition (\ref{eq:IntCond}) is satisfied with $\O(g)=-g^{-1}$ on the boundary.
\end{quote}
\label{table:AdSSTDbranes}
\end{table}
\normalsize

Again we summarizes the possible $U$'s
\be
U=\bigg\{
\left(
  \begin{array}{cc}
    P_{a}/P_{0 1 2 3 4 } & {} \\
    {} & P_{a'b'}/P_{\tilde a'\tilde b'} \\
  \end{array}
\right),
\left(
  \begin{array}{cc}
    P_{\tilde a} & {} \\
    {} & P_{\tilde a'b'}\\
  \end{array}
\right),
\ee
$$
~~~~~~~~~~~~\left(
  \begin{array}{cc}
    P_{a b}/P_{\tilde a \tilde b} & {} \\
    {} & P_{a'}/P_{5,6,7,8,9} \\
  \end{array}
\right),
\left(
  \begin{array}{cc}
    P_{a \tilde b} & {} \\
    {} & P_{\tilde a'}\\
  \end{array}
\right)
\bigg\},
$$
where $a,b=0,1,3$, $\tilde a,\tilde b=2,4$, $a',b'=5,6,8$ and $\tilde a',\tilde b'=7,9$. We do not give the relative coefficient between the two block. $P_a,P_{\tilde a},P_{ab},P_{\tilde a\tilde b},P_{a\tilde b}$ and $P_{1,2,3,4,5}$ gives 2, 4, 4, 0, 2 and 2-dimensional Neumann boundary conditions respectively and similarly for the $P$'s with the primed indices.

\section{Integrable configurations for the $\mathrm{AdS}_4\times \mathbb{C}\mathrm{P}^3$ Background}\label{sec:Integrable configurations for the AdS4xCP3}
As for the $\mathrm{AdS}_5\times \mathrm{S}^5$ case we can analyze the $\mathrm{AdS}_4\times \mathbb{C}\mathrm{P}^3$ background and find integrable configurations.
The $\mathrm{AdS}_4\times \mathbb{C}\mathrm{P}^3$ background is constructed using the supercoset $\frac{\mathrm{OSP}(6|4)}{\mathrm{U}(3)\times \mathrm{SO}(3,1)}$, which has $\mathbb{Z}_4$ grading structure and so is an integrable background, see \cite{Arutyunov:2008if}\cite{Stefanski:2008ik}.
First we present general results for this background for the bosonic sector and then we examine the $\mathrm{AdS}_3\times \mathbb{C}\mathrm{P}^1$ configuration.
Finally, we carry out a more general analysis and find more examples of integrable configurations.

The relevant superalgebra is OSP$(6|4)$, the details are found in appendix \ref{ap:OSP64SCA}. The outer automorphism for the OSP$(2k|2n)$ superalgebra is \cite{1985IzMat..24..539S}
$\mathrm{Ad}J_{k,n}$, with $J_{k,n}\in \mathrm{GL}(2k,2n)$,
$\det J_{k,n}=-1$, $J^2_{k,n}=1_{2k+2n}$ and
$J_{k,n}B_{2k,n}J_{k,n}=B_{2k,n}$ where
$B_{2k,n}=\mathrm{diag}(1_{2k},J_n)$,
$J_n=\tiny\left(
                                     \begin{array}{cc}
                                       0 & 1_n \\
                                       -1_n & 0 \\
                                     \end{array}
                                   \right)$.
So we consider the automorphism generally acting as $\O(x)=U x U^{-1}$ with $U\in \mathrm{GL}(6|4)_{\bar 0}$.

\subsection{The $\mathrm{AdS}_4\times \mathbb{C}\mathrm{P}^3$ bosonic background in global coordinates}
We write the $\mathrm{AdS}_4\times \mathbb{C}\mathrm{P}^3$ metric in global coordinates
as $ds^2=ds^2_{\mathrm{AdS}_4}+4ds^2_{\mathbb{C}\mathrm{P}^3}$ with
\be
ds^2_{\mathrm{AdS}_4}=d\r^2
-\cosh^2\r dt^2
+\sinh^2\r (d\a^2
+\sin^2\a d\b^2)
\ee
\be
ds^2_{\mathbb{C}\mathrm{P}^3}=d\m^2+\cos^2\m\sin^2\m(d\psi-\frac{1}{2}\cos\t_1 d\varphi_1+\frac{1}{2}\cos\t_2 d\varphi_2)^2
\ee
$$
+\frac{1}{4}\sin^2\m(d\t_1^2+\sin^2\t_1 d\varphi_1^2)
+\frac{1}{4}\cos^2\m(d\t_2^2+\sin^2\t_2 d\varphi_2^2).
$$
In order to get this form of the metric we take the coset representative to be
\be
g_{\mathrm{AdS}_4}=e^{P_0 t}e^{J_{12}\b}e^{J_{13}\a}e^{P_3 \r}
\ee
\be
g_{\mathbb{C}\mathrm{P}^3}=e^{M_{25}\psi}e^{-R_4\varphi_2}e^{R_3 (\t_2+\frac{\pi}{2})}e^{T_4\varphi_1}e^{T_3 (\t_1+\frac{\pi}{2})}e^{2R_6\m}
\ee
where the generators are defined in appendix \ref{ap:OSP64SCA}.

Then the bosonic sector's $J^{(2)}$ is given by
\be
J^{(2)}=
-P_0\cosh\r dt
-P_1\sinh\r d\a
-P_2\sinh\r \sin\a d\b
+P_3 d\r
\ee
$$
+R_1\sin\m d \t_1
+R_2\sin\m\sin\t_1 d \varphi_1
+R_3\cos\m d \t_2
+R_4\cos\m\sin\t_2 d\varphi_2
$$
$$
-R_5\cos\m\sin\m(\cos\t_1 d\varphi_1-\cos\t_2 d\varphi_2-2 d\psi)
+2R_6 d\m.
$$

In order for the transformed generator $\O(x)$, $x\in \mathfrak{g}$ to be in the superalgebra $U$ must satisfy
\be\label{eq:OSP_U_cond}
U^t=\pm B U^{-1} B^{-1},\quad
B=1_6\oplus J_4,
\ee
see appendix \ref{ap:OSP64SCA} for notations.

\subsection{Karch-Randall D4-Brane}
In this case the D4-brane has the topology $\mathrm{AdS}_3\times \mathbb{C}\mathrm{P}^1$. We take the boundary conditions $\b=0$ and $\m=0$.
Then we find that
\be
J^{(2)}=
-P_0\cosh\r dt
-P_1\sinh\r d\a
-P_2\sinh\r \sin\a d\b
+P_3 d\r
\ee
$$
+R_3d \t_2
+R_4\sin\t_2 d\varphi_2
+2R_6 d\m,
$$
and in worldsheet components
\be
J^{(2)}_\tau=
-P_0\cosh\r \p_\tau t
-P_1\sinh\r \p_\tau \a
+P_3 \p_\tau \r
+R_3\p_\tau  \t_2
+R_4\sin\t_2 \p_\tau \varphi_2,
\ee
\be
J^{(2)}_\s=
-P_2\sinh\r \sin\a \p_\s\b
+2R_6 \p_\s\m.
\ee
The matrix that satisfies (\ref{eq:integrabilitycondition1}) is
\be
U=a P_0 P_1 P_3+ b (M_{14}^2-M_{25}^2+M_{36}^2),\quad
a,b\in \mathbb{C}.
\ee
This $U$ also satisfies $[U,g]=0$ at the boundary. In order for this automorphism to be involutive we must have $b=\pm a$.
We further need the the transformed generators $\O(x)$, $x\in \mathfrak{g}$ to be in the superalgebra so $U$ must satisfy
\be\label{eq:OSP_U_cond}
U^t=\pm B U^{-1} B^{-1},\quad
B=1_6\oplus J_4,
\ee
see appendix \ref{ap:OSP64SCA} for notations.
The relation $a=\pm b$ can be checked to be consistent with (\ref{eq:OSP_U_cond}).
We fix $U=8 P_0 P_1 P_3 + (M_{14}^2-M_{25}^2+M_{36}^2)$ so that we have a solution to (\ref{eq:integrabilitycondition2}), with
\be
U Q^1_{\a a'} U^{-1}=(U_A)_\a{}^\b (U_{CP})_{a'}{}^{b'}Q^3_{\b b'},\quad
U Q^3_{\a a'} U^{-1}=(U_A)_\a{}^\b (U_{CP})_{a'}{}^{b'}Q^1_{\b b'},
\ee
where we defined
\be
U=\left(
    \begin{array}{cc}
      U_{CP} & 0 \\
      0 & U_A \\
    \end{array}
  \right).
\ee
Note there is no relative minus sign for the $Q^1$ and $Q^3$ transformations as was in the $\mathrm{AdS}_5\times \mathrm{S}^5$ case.
Similarly to the $\mathrm{AdS}_5\times \mathrm{S}^5$ case we define $F=Q^I_{\a a'} \t^{I \a a'}$.
We require $F$ to be invariant under the $U$ conjugation, and so impose the boundary conditions
\be
(U_{A})_\a{}^\b(U_{CP})_{a'}{}^{b'}\t^{1\a a'}=\t^{3\b b'},\quad
(U_{A})_\a{}^\b(U_{CP})_{a'}{}^{b'}\t^{3\a a'}=\t^{1\b b'}.
\ee
Then (\ref{eq:integrabilitycondition2}) is satisfied.
We conclude that the D-brane configuration is integrable, preserving half of the supersymmetry.

\subsection{Other integrable configurations}
Similarly to the $AdS_5\mathrm{\times} \mathrm{S}^5$ we can find other integrable D-brane configurations repeating the analysis above.
We note that at the bosonic level we can always find $U$ which commutes with any set of $P$'s and anti-commute with the rest.

In table \ref{table:AdS4Dbranes} we give examples of integrable D-brane configurations.
Our analysis for the $\mathrm{AdS}_4\times \mathbb{C}\mathrm{P}^3$ background is less systematic then the one we gave for the $AdS_5\mathrm{\times} \mathrm{S}^5$ background.
Generally, we have automorphisms that work as conjugation with some invertible matrix $U$.
We require equations (\ref{eq:integrabilitycondition1}-\ref{eq:integrabilitycondition3}) to be satisfied with $\O$ being an involutive automorphism.
We further require (\ref{eq:OSP_U_cond}) to be satisfied so that the transformed generators will stay in the superalgebra.
We do not know of a complete classification of half-BPS D-branes in this background, see \cite{Chandrasekhar:2009ey} for some results.

\begin{table}[ht]
\caption{Integrable half-BPS D-brane configurations for $\mathrm{AdS}_4\times \mathbb{C}\mathrm{P}^3$ with $\O(x)=U x U^{-1}$.}
\centering
\scriptsize
\begin{tabular}{|l ||l| l|}
\hline\hline
$\mathbb{R}^{1,0} $                               &$U=P_{0}+(M_{14}+M_{25}+M_{36})$ &$\r=0;\m=0,\t_2=-\frac{\pi}{2},\varphi_2=0$\\
$\mathbb{R}^{0,1}_+ $                             &$U=P_{3}+i(M_{14}+M_{25}+M_{36})$ &$t=0,\a=0;\m=0,\t_2=-\frac{\pi}{2},\varphi_2=0$\\
$\mathrm{S}^3\subset \mathbb{C}\mathrm{P}^3$      &$U=P_{0,1,2,3}+2i(T_1 R_1+T_3 R_3+T_6 R_6)$ &$\r=0,t=0;\varphi_1=\varphi_2=\psi=0$\\
$\mathrm{AdS}_2\times \mathrm{S}^3$               &$U=P_{0,3}+2i(T_1 R_1+T_3 R_3+T_6 R_6)$ &$\a=0;\varphi_1=\varphi_2=\psi=0$\\
$\mathrm{AdS}_3\times \mathrm{S}^2$               &$U=P_{0,1,3}+(M_{14}^2-M_{25}^2+M_{36}^2)$ &$\b=0,\m=0$\\
$\mathbb{H}^3\times \mathrm{S}^2$                 &$U=P_{1,2,3}+i(M_{14}^2-M_{25}^2+M_{36}^2)$ &$t=0,\m=0$\\
$\mathrm{AdS}_4\times \mathrm{S}^3$               &$U=P_{1,1}+2(T_1 R_1+T_3 R_3+T_6 R_6)$ &$\varphi_1=\varphi_2=\psi=0$\\
$\mathrm{AdS}_3\times \mathbb{C}\mathrm{P}^3$     &$U=P_{0,1,3}+(M_{14}^2+M_{25}^2+M_{36}^2)$ &$\b=0$\\
$\mathbb{H}^3\times \mathbb{C}\mathrm{P}^3$       &$U=P_{1,2,3}+i(M_{14}^2+M_{25}^2+M_{36}^2)$ &$t=0$\\
\hline
\end{tabular}
\begin{quote}
We give some of the integrable D-brane configurations with the gluing conditions involving $\O(x)=U x U^{-1}$ automorphism.
The second column gives the $U$ matrices in terms of the superalgebra generators. The notation is such that $P_{a,b,c,...}=(2P_a)(2P_b)(2P_c)...$. For each boundary condition (\ref{eq:IntCond}) is satisfied with $\O(g)=g$ on the boundary.
\end{quote}
\label{table:AdS4Dbranes}
\end{table}
\normalsize

\section{Discussion}\label{sec:Discussion}
In this paper we introduced a procedure for constructing a generating function for an infinite set of conserved charges for the GS$\s$M with boundaries, by generalizing
methods that were used for $(1+1)$-dimensional bosonic sigma-models.
We considered the full sector including the fermionic degrees of freedom, and found a class of boundary conditions that break half of the supersymmetry. The boundary conditions are expressed using the simple equation
\be
a(z)=\O(\bar a(z^{-1})),
\ee
where $a$ is the flat connection and $\O$ is an involutive metric preserving automorphism.
We found that these boundary conditions imply integrability of the boundary configuration.

We constructed some general solutions for the automorphism maps $\O$ for the $\mathrm{AdS}_5\times \mathrm{S}^5$ and $\mathrm{AdS}_4\times \mathbb{C}\mathrm{P}^3$ backgrounds, and gave examples of integrable configurations in both cases, see tables \ref{table:AdSnoSTDbranes}, \ref{table:AdSSTDbranes} and \ref{table:AdS4Dbranes}.
Among these examples we found the $\mathrm{AdS}_4\times \mathrm{S}^2$ configuration to be integrable, which was recently claimed to be so, see \cite{Correa:2011nz},\cite{MacKay:2011zs},\cite{Regelskis:2011fa}.

Our analysis of integrability is classical,
it will be interesting to find whether the integrable structure that we found survives quantization.

The integrable D-brane configurations we constructed are half-BPS.
It is interesting to check, whether there are D-brane configurations that breaks more than one-half of the supersymmetries and are still integrable.
As for flat-space, we can find open strings stretching between half-BPS D-brane configurations that will preserve some of the supersymmetry, e.g a string stretching between $\mathbb{R}^1\times \mathrm{S}^3$ and $\mathrm{AdS}_5\times \mathrm{S}^3$ (with the same $\mathrm{S}^3$) preserves 1/4 of the supersymmetries. Using our construction these configurations should also be integrable.

The gluing conditions that we used are not the most general ones for which the boundary term vanishes and the action is invariant.
Nonetheless, if we are interested in integrable configurations it is plausible that the automorphism should relate the flat-connection to some other flat-connection as in (\ref{eq:BC}) (possibly with a different function of the spectral parameter), which seems to be satisfied only for the gluing conditions that we used.
The gluing condition (\ref{eq:MovingFCTrans}) with $z^{-1}$ replaced with $(iz)^{-1}$ seems to be as a good candidate, but the analog of (\ref{eq:BC}) is not satisfied.

It will be interesting to consider the gauge dual operators corresponding to the integrable D-brane configurations that we found, and compare the integrable structure.

\section*{Acknowledgements}
We would like to thank Ido Adam and Sunny Itzhaki for valuable discussions.
The work is supported in part by the Israeli
Science Foundation center of excellence,  by the US-Israel Binational Science
Foundation (BSF), and by the German-Israeli Foundation (GIF).

\appendix
\section{The PSU$(2,2|4)$ superconformal algebra}\label{ap:PSU224SCA}
The PSU$(2,2|4)$ superconformal algebra in the $\mathrm{so}(4,1)\otimes \mathrm{so}(5)$ basis, which will be convenient for treating the background in global coordinates is given by\footnote{This superalgebra agrees with the one in \cite{Metsaev:1998it} up to normalization of the odd generators by factor of $\sqrt{-2i}$ and $P\rightarrow -P$.}
\be
[P_a,P_b]=J_{ab},\quad
[P_{a'},P_{b'}]=-J_{a'b'}
\ee
$$
[P_a,J_{bc}]=\eta_{ab}P_{c}-\eta_{ac}P_{b},\quad
[P_{a'},J_{b'c'}]=\eta_{a'b'}P_{c'}-\eta_{a'c'}P_{b'},\quad
$$
$$
[J_{ab},J_{cd}]=\eta_{bc}J_{ad}+~\mathrm{perm},\quad
[J_{a'b'},J_{c'd'}]=\eta_{b'c'}J_{a'd'}+~\mathrm{perm},\quad
$$
$$
[Q^I_{\a\a'},P_{a}]=\frac{i}{2}\e^{IJ}Q^J_{\b\a'}(\g_a)_\a{}^\b,\quad
[Q^I_{\a\a'},P_{a'}]=-\frac{1}{2}\e^{IJ}Q^J_{\a\b'}(\g_{a'})_{\a'}{}^{\b'},\quad
$$
$$
[Q^I_{\a\a'},J_{ab}]=-\frac{1}{2}Q^I_{\b\a'}(\g_{ab})_\a{}^\b,\quad
[Q^I_{\a\a'},J_{a'b'}]=-\frac{1}{2}Q^I_{\a\b'}(\g_{a'b'})_{\a'}{}^{\b'},\quad
$$
$$
\{Q^{I}_{\a\a'},Q^{J}_{\b\b'}\}
=\d^{IJ}\left(C'_{\a'\b'}(C\g^{a})_{\a\b}P_{a}+iC_{\a\b}(C'\g^{a'})_{\a'\b'}P_{a'}+C'_{\a'\b'}C_{\a\b}A\right)
$$
$$
-i\e^{IJ}\frac{1}{2}\left(C'_{\a'\b'}(C\g^{ab})_{\a\b}J_{ab}-C_{\a\b}(C'\g^{a'b'})_{\a'\b'}J_{a'b'}\right)
$$
with $a=0,..,4$, $a'=5,..,9$, $\a=1,2$, $\a'=1,2$, $I=1,2$.
The $32\times 32$ gamma matrices are given by $\Gamma_a=\g_a\otimes 1\otimes \s_1$, $\Gamma_{a'}=1 \otimes \g_{a'}\otimes \s_2$ with
\be\label{eq:smallgammas}
\g_0=i\s_3\otimes 1,\quad
\g_1=\s_2\otimes\s_2,\quad
\g_2=-\s_2\otimes\s_1,\quad
\g_3=\s_1\otimes 1,\quad
\g_4=\s_2\otimes\s_3,\quad
\ee
$$
\g_5=\s_3\otimes 1,\quad
\g_6=\s_2\otimes\s_2,\quad
\g_7=\s_2\otimes\s_1,\quad
\g_8=\s_1\otimes 1,\quad
\g_9=-\s_2\otimes\s_3,\quad
$$
and the charge conjugation matrix $\mathcal{C}=C\otimes C\otimes i\s_2$ with $C_{\a\b}=(\g_2\g_4)_\a{}^\b=i 1\otimes \s_2$ and
$C'_{\a'\b'}=(\g_7\g_9)_{\a'}{}^{\b'}=i 1\otimes \s_2$ the charge conjugation matrices of the so(4,1) and so(5) spinors respectively ($C\g^{a}C^{-1}=\g^{a t}$, $a=0,..,4$).
We normalize $\g^{ab}=\frac{1}{2}[\g^a,\g^b]$.
$A$ is the U$(1)$ factor of the $\mathrm{SU}(2,2|4)$ superconformal-algebra, which we drop in order to get $\mathrm{PSU}(2,2|4)$.
The charge conjugation matrix acts on the gamma matrices in the standard way, $C \G^{\hat a} C^{-1}=-(\G^{\hat a})^t$.
The odd matrices satisfy $(Q^{I}_{\a\a'})^\dag=C^{\a\b}C^{\a'\b'}Q^I_{\b\b'}$, or in the super-matrix algebra basis $(Q^{I}_{\a\a'})^\dag=k Q^I_{\a\a'}k^{-1}$ with $k=\tiny\left(
    \begin{array}{cc}
      \g^2\g^4 & 0 \\
      0 & \g^7\g^9 \\
    \end{array}
  \right)$.

The $\mathbb{Z}_4$ automorphism map is given by
\be
\hat\O(X)=-K X^{st} K^{-1},\quad
K=\left(
    \begin{array}{cc}
      \g^2\g^4 & 0 \\
      0 & i\g^7\g^9 \\
    \end{array}
  \right)
  =\left(
    \begin{array}{cc}
      i\otimes \s_2 & 0 \\
      0 &  -1 \otimes \s_2 \\
    \end{array}
  \right),
\ee
where
\be
\left(
    \begin{array}{cc}
      A & B \\
      C & D \\
    \end{array}
  \right)^{st}
  =
\left(
    \begin{array}{cc}
      A^t & C^t \\
      -B^t & D^t \\
    \end{array}
  \right).
\ee
The $\mathbb{Z}_4$ grading subspaces are spanned by
\be
\mathcal{H}_0=\{J_{\hat a \hat b}\},\quad
\mathcal{H}_1=\{Q^1_{\a\a'}\},\quad
\mathcal{H}_2=\{P_{\hat a}\},\quad
\mathcal{H}_3=\{Q^2_{\a\a'}\}.
\ee

\section{The OSP$(6|4)$ superconformal algebra}\label{ap:OSP64SCA}
The bosonic sp(4) subalgebra is given by
\be
[P_a,P_b]=J_{ab},\quad
[P_a,J_{bc}]=\eta_{ab}P_{c}-\eta_{ac}P_{b},\quad
[J_{ab},J_{cd}]=\eta_{bc}J_{ad}+~\mathrm{perm},\quad
\ee
with $a,b=0,..,3$, with $\eta=(-,+,+,+)$.
The so(6) algebra is given by
\be
[M_{a'b'},M_{c'd'}]=\d_{b'c'}M_{a'd'}+~\mathrm{perm},\quad
\ee
with $a',b'=1,..,6$.
The commutation with the odd generators is given by
\be
[M_{a'  b'}, Q_{\a c'}]=\d_{a'c'}Q_{\a b'}-\d_{b'c'}Q_{\a a'},
\ee
\be
[P_{a}, Q_{\a c'}]=-\frac{1}{2}Q_{\b c'}(\g_{a})_\a{}^\b,\quad
[J_{a b}, Q_{\a c'}]=-\frac{1}{2}Q_{\b c'}(\g_{ab})_\a{}^\b,
\ee
\be
\{Q_{\a a'},Q_{\b b'}\}=\d_{a'b'}(P_a(\g^a C)_{\a\b}-\frac{1}{2}J_{ab}(\g^{ab}C)_{\a\b})-C_{\a\b}M_{a'b'},
\ee
with $\a=1,..,4$. The gamma matrices are given by
\be
(\g^0)_\a{}^\b=i\otimes \s^2,\quad
(\g^1)_\a{}^\b=\s^3\otimes \s^1,\quad
(\g^2)_\a{}^\b=\s^1\otimes \s^1,\quad
(\g^3)_\a{}^\b=-\s^2\otimes \s^1,
\ee
and the C.C matrix is given by $C_{\a\b}=i\s^2\otimes 1$.

The $\mathbb{Z}_4$ automorphism map that gives $\mathrm{AdS}_4\times \mathbb{C}\mathrm{P}^3$ is given by \cite{Serganova83}
\be
\hat\O(x)=K x K^{-1}, \quad
K=J_6\oplus I^1_2 \oplus I^1_2,\quad
J_{2k}=\left(
         \begin{array}{cc}
           0 & 1_k \\
           -1_k & 0 \\
         \end{array}
       \right),\quad
I^l_n=\mathrm{diag}(1_l,-1_{n-l}).
\ee
The automorphism decomposes the so(6) algebra with respect to its u(3) subalgebra. The coset graded-2 generators are\footnote{A similar decomposition is given in \cite{Girardi:1980um}.}
\be
R_1 = \half(M_{1, 2} - M_{4, 5}),\quad
R_2 = \half(M_{1, 5} - M_{2, 4}),\quad
R_3 = \half(M_{1, 3} - M_{4, 6}),\quad
\ee
$$
R_4 = \half(M_{1, 6} - M_{3, 4}),\quad
R_5 = \half(M_{2, 6} - M_{3, 5}),\quad
R_6 = \half(M_{2, 3} - M_{5, 6}),\quad
$$
and the graded-0 u$(3)$ generators are
\be
T_1 = \half(M_{1, 2} - M_{4, 5}),\quad
T_2 = \half(M_{1, 5} - M_{2, 4}),\quad
T_3 = \half(M_{1, 3} - M_{4, 6}),\quad
\ee
$$
T_4 = \half(M_{1, 6} - M_{3, 4}),\quad
T_5 = \half(M_{2, 6} - M_{3, 5}),\quad
T_6 = \half(M_{2, 3} - M_{5, 6}),\quad
$$
$$
T_7=M_{1,4},\quad
T_8=M_{2,5},\quad
T_9=M_{3,6}.
$$

We note that
\be
\hat\O(Q_{\a a'})=-(I^1_2 \oplus I^1_2)_\a{}^\b (J_6)_{a'}{}^{b'} Q_{\b b'},
\ee
so
\be
Q^{(1)/(3)}_{\a a'}=\frac{1}{2}(\d_\a^\b \d_{a'}^{b'}\mp i (I^1_2 \oplus I^1_2)_\a{}^\b (J_6)_{a'}{}^{b'} )Q_{\b b'}
\ee
for $\a=1,..,4$, $a'=1,2,3$.

\section{More parameterizations for the AdS$_5\times $S$^5$ coset}\label{ap:more_parameterizations}
In the main text we gave a parametrization for the AdS$_5\times $S$^5$ background in global coordinates for the AdS and Hopf coordinates for the sphere. Here we give a parametrization for the AdS subspace in Poincar\'{e} coordinates. We use the parametrization
\be
g=\exp(p_\m x^\m)y^D
\ee
where
\be
p_\m=P_{\m}-J[\m,5],\quad
D=P_4.
\ee
The metric in this case is
\be
ds^2_{\mathrm{AdS}_5}=\frac{dx^\m dx_\m+dy^2}{y^2}
\ee
with $\eta=\mathrm{diag}(-,+,+,+)$.
The current $J^{(2)}$ for the bosonic sector is
\be
J^{(2)}_{\mathrm{AdS}_5}=\frac{1}{y}(P_0 dt+P_1 dx_1+P_2 dx_2+P_3 dx_3-P_4 dy).
\ee
The sphere's metric can be written in the usual spherical coordinates using the parametrization
\be
g=\prod_{a'=5}^{9}\exp(P_{a'}\t_{a'})=\exp(P_{5}\t_{5})\exp(P_{6}\t_{6})...
\ee
so that
\be
ds^2_{\mathrm{S}^5}=d\t_{9}^2+\cos^2\t_{9}(d\t_{8}^2+\cos^2\t_{8}(d\t_{7}^2+\cos^2\t_{7}(d\t_{6}^2+\cos^2\t_{6}(d\t_5^2+\cos^2\t_5))).
\ee
The current $J^{(2)}$ for the bosonic sector is
\be
J^{(2)}_{\mathrm{S}^5}=P_{9}d\t_{9}+\cos\t_{9}(P_{8}d\t_{8}+\cos\t_{8}(P_{7}d\t_{7}+\cos\t_{7}(P_{6}d\t_{6}+\cos\t_6 P_{5}d\t_5))).
\ee

\bibliography{postdoc}
\bibliographystyle{JHEP}
\end{document}